\documentclass{article}

 \usepackage{graphicx}% Include figure files
\usepackage{bm}% bold math
\usepackage{amsfonts}
\usepackage{amsmath}

\usepackage{graphicx}
\usepackage{epsfig}
\usepackage{amsfonts}
\usepackage{a4}
\usepackage{amsmath}

\def\dalemb#1#2{{\vbox{\hrule height .#2pt
        \hbox{\vrule width.#2pt height#1pt \kern#1pt
                \vrule width.#2pt}
        \hrule height.#2pt}}}

\def\0{{\sst{(0)}}}
\def\1{{\sst{(1)}}}
\def\2{{\sst{(2)}}}
\def\3{{\sst{(3)}}}
\def\4{{\sst{(4)}}}
\def\5{{\sst{(5)}}}
\def\6{{\sst{(6)}}}
\def\7{{\sst{(7)}}}
\def\8{{\sst{(8)}}}

\def\ep{\epsilon}

 \def\bd{\begin{document}} \def\ed{\end{document}}
\def\ds{\documentstyle} \let\fr=\frac \let\bl=\bigl \let\br=\bigr
\let\Br=\Bigr \let\Bl=\Bigl
\let\bm=\bibitem
\let\na=\nabla
\let\pa=\partial \let\ov=\overline
\def\fft#1#2{{#1 \over #2}}

\def\oneone{\rlap 1\mkern4mu{\rm l}}
\def\ep{{\epsilon}}

\begin{document}

\title{
%Phase transitions 
Thermodynamic properties of asymptotically anti-de Sitter
black holes
 in $d=4$ Einstein--Yang-Mills theory 
 %with negative cosmological constant
 }

\author{{\large Olga Kichakova}$^{1}$, {\large  Jutta Kunz}$^{1}$,
{\large Eugen Radu}$^{2}$
and {\large Yasha Shnir}$^{1,3,4}$
\\
\\
 {\small  $^{1}$Institut f\"ur Physik, Universit\"at Oldenburg, Postfach 2503
D-26111 Oldenburg, Germany}
\\
 {\small  $^{2}$Departamento de Fisica da Universidade de Aveiro and I3N}
 \\
 {\small  Campus de Santiago, 3810-183 Aveiro, Portugal}
\\
{\small $^{3}$Department of Theoretical Physics, Tomsk State Pedagogical University,
Russia}
\\
{\small $^{4}$BLTP, JINR, Dubna, Russia}
 }
\date{\today}

\maketitle

\begin{abstract} 
We investigate the thermodynamics of
spherically symmetric black hole solutions in a four-dimensional Einstein--Yang-Mills-SU(2) theory with a negative cosmological constant.  
Special attention is paid to configurations with a unit magnetic charge.
We find that a set of Reissner-Nordstr\"om--Anti-de Sitter black holes
can become unstable to forming non-Abelian hair.
However, the hairy black holes are never 
thermodynamically favoured over the full set of abelian monopole solutions.
The thermodynamics of the generic configurations possessing a noninteger 
magnetic charge is also discussed.

\end{abstract}

%%%%%%%%%%%%%%%%%%%%%%%%%%%%%%%%%%%%%%%%%%%%%%%%%%%%%%%%%%%%%%%%%%%%%%%%%%%%%%
\section{Introduction}
%%%%%%%%%%%%%%%%%%%%%%%%%%%%%%%%%%%%%%%%%%%%%%%%%%%%%%%%%%%%%%%%%%%%%%%%%%%%%%
 
Black holes are non-perturbative objects whose existence appears to be an unavoidable
consequence of general relativity (and its various extensions).
Moreover, as often stated in the literature,
the black holes (BHs) are the quantum gravity counterparts of
the hydrogen atom in ordinary quantum mechanics. 
Thus some basic results derived at the semiclassical level, like the existence of 
Hawking radiation together with an intrinsic BH entropy 
are expected to be very basic features that any putative quantum theory
of gravity will have to take into account.

As a result, the subject of BH thermodynamics 
has enjoyed a constant interest over the last four decades.
BH solutions in anti-de Sitter (AdS) spacetime background
have been considered also in this context.
For example, as shown by Hawking and Page \cite{Hawking:1982dh}, the presence of
a negative comological constant  makes it possible for a BH to reach stable thermal equilibrium
with a heat bath.
Moreover, according to the AdS/CFT conjecture 
\cite{Maldacena:1997re},
%\cite{Witten:1998qj},
 BH solutions with AdS asymptotics would offer the possibility of probing
the nonperturbative structure of some conformal field theories. 
%For example, the AdS$_5$ Hawking-Page
%phase transition  is interpreted as a thermal phase transition from a confining to a deconfining phase in
%the dual $d = 4, {\cal N} = 4$ super Yang-Mills theory 
%\cite{Witten:1998zw}, 
%while the phase structure of Reissner- Nordstr\"om-AdS (RNAdS)
%black holes resembles that of a van der Waals-Maxwell liquid-gas system
%\cite{Chamblin:1999tk},
%\cite{Chamblin:1999hg}.

The study of thermodynamics of BH
solutions violating the no hair conjecture
is particularly interesting. 
This conjecture states that the
only allowed characteristics of a stationary BH are those associated with the Gauss law, such as
mass, angular momentum and U(1) charges \cite{Wheeler}. 
Apart from a pure mathematical interest, 
the BHs with hair may be useful for probing not only
quantum gravity, but also may 
 play an important role in the 
context of
the AdS/CFT correspondence.

%be interesting tests of the AdS/CFT correspondence.

The first (and still the best known) example of hairy BH solutions are those
in Einstein--Yang-Mills (EYM) theory.
Moreover, this example can be regarded as canonical
in the sense that other hairy solutions 
usually share a number of common characteristics
with the EYM case 
(a review of hairy non-Abelian BHs solutions with a cosmological constant $\Lambda \geq 0$
can be found in \cite{Volkov:1998cc}).
BHs  with non-Abelian (nA) hair in AdS background have also been
extensively studied, starting with the pioneering work 
\cite{Winstanley:1998sn}.
These EYM solutions possess a variety of interesting features which strongly contrast
with those of the asymptotically flat spacetime counterparts in  \cite{Volkov:1989fi}.
For example, stable BHs with a  magnetic charge
are known to exist even in the absence of a Higgs field  
(see 
\cite{Winstanley:2008ac}
for  a review of these solutions).  

Moreover,
considering such configurations is a legitimate task, since the gauged supergravity
models
(of interest in AdS/CFT context) generically contain the EYM action as the basic buiding block.

The main purpose of this work is to address the
issue of the thermodynamical behaviour
of the AdS   solutions with a spherical
horizon topology
in EYM-SU(2) theory,
a problem which, to our knowledge, has not yet been addressed in a systematic way.
Special attention is paid to configurations sharing the asymptotics
with the Schwarzschild-AdS and the (embedded Abelian) Reissner-Nordstr\"om-AdS BHs.
Also, for simplicity, we shall restrict our study to
configurations featuring magnetic fields only.

Our results show that the EYM solutions possess a variety of new
features,  
the generic picture depending on 
the value of the magnetic charge
as well as on the ratio of the four-dimensional gravitational
constant to the Yang-Mills coupling.
Moreover, as we shall see, the thermodynamical properties
of the solutions
depend also on the topology of the horizon, the case of spherical
configurations being special.

%%%%%%%%%%%%%%%%%%%%%%%%%%%%%%%%%%%%%%%%%%%%%%%%%%%%%%%%%%%%%%%%%%%%%%%%%%%%%%
\section{The solutions}
%%%%%%%%%%%%%%%%%%%%%%%%%%%%%%%%%%%%%%%%%%%%%%%%%%%%%%%%%%%%%%%%%%%%%%%%%%%%%%

The hairy BHs discussed in this work are solutions of the Einstein--Yang-Mills-SU(2) equations with a negative
 cosmological constant $\Lambda=-3/L^2$,
\begin{eqnarray}
%\nonumber
\label{EYM-eqs}
&&R_{\mu\nu}-\frac{1}{2}R g_{\mu\nu}-\frac{3}{L^2} g_{\mu\nu}=8\pi G 
\left(
F_{\mu \alpha}^{(a)} F_{\nu \beta}^{(a)} g^{\alpha\beta}-\frac{1}{4}g_{\mu\nu} F_{\alpha \beta}^{(a)}F^{(a)\alpha \beta}
\right),
%\\
%&&
~~D_\mu F^{\mu\nu}=0.
\end{eqnarray}
 
%\begin{eqnarray}
%S=\int_{\cal M}d^4 x 
%\left (
%R+\frac{6}{L^2}-\frac{1}{4}F_{\mu\nu}F^{\mu\nu}
%\right)
%\end{eqnarray}

We are mainly interested in spherically symmetric solutions,
with magnetic fields only, a case
which can be studied within the following ansatz
\begin{eqnarray}
\label{metric-ansatz}
ds^2=\frac{dr^2}{N(r)}+r^2(d\theta^2+\sin^2 \theta d\phi^2)-\sigma^2(r)N(r) dt^2,~~~
{\rm with}~~~N(r)=1-\frac{2m(r)}{r}+\frac{r^2}{L^2},
\end{eqnarray}
for the metric, and
\begin{eqnarray}
\label{YM-ansatz}
A=\frac{1}{2\hat g} \big [
w(r) \tau_1 d\theta+
(
\cos \theta \tau_3 + w(r)\sin \theta \tau_2 
)
d\phi
\big ],
\end{eqnarray}
for the gauge fields (with $\hat g$ the gauge coupling constant and $\tau_a$ the Pauli matrices).
Then the field equations (\ref{EYM-eqs}) reduce to 
\begin{eqnarray}
\label{eqs1}
m'=\alpha^2 \left(
%Nw'^2+\frac{(w^2-1)^2}{2r^2}
Nw'^2+\frac{V^2(w)}{2r^2}
\right),~~
\sigma'=2\alpha^2\sigma \frac{w'^2}{r},~~
%\\
%\label{eqs2}
%&&
%w''+\left( \frac{N'}{N}+\frac{\sigma'}{\sigma} \right)w'+\frac{w(1-w^2)}{r^2N}=0.
w''+\left( \frac{N'}{N}+\frac{\sigma'}{\sigma} \right)w'+\frac{wV(w)}{r^2N}=0,
\end{eqnarray}
(with  $V(w)=(1-w^2)$),
where we have defined the coupling constant
\begin{eqnarray}
\alpha^2=\frac{4\pi G}{\hat g^2}.
\end{eqnarray}

We want the metric (\ref{metric-ansatz}) to describe a nonsingular, asymptotically AdS spacetime outside a
 horizon located
at $r = r_H>0$  
(here $N(r_H) = 0$ is only a coordinate singularity where all curvature invariants are finite).
There are two explicit solutions satisfying these assumptions.
If $w^2(r)\equiv  1$, then 
\begin{eqnarray}
\label{SAdS}
N(r)=\left(1-\frac{r_H}{r}\right)
\left(
1+\frac{r^2}{L^2}+\frac{r r_H}{L^2}+\frac{r_H^2}{L^2}
\right),~~\sigma(r)=1,
\end{eqnarray}
which is just the Schwarzschild-AdS (SAdS) metric, with vanishing YM curvature and mass $M=\frac{r_H}{2}(1+\frac{r_H^2}{L^2})$.
A different solution is found for  $w(r) \equiv 0$, with
\begin{eqnarray}
\label{RNAdS}
N(r)=\left(1-\frac{r_H}{r}\right)
\left(
1+\frac{r^2}{L^2}+\frac{r r_H}{L^2}+\frac{r_H^2}{L^2}
-\frac{\alpha^2}{r r_H}
\right), ~~\sigma(r)=1.
\end{eqnarray}
This describe the embedded Abelian magnetic Reissner-Nordstr\"om--AdS (RNAdS) metric, 
with mass $M=\frac{1}{2r_H}(\alpha^2+r_H^2(1+\frac{r_H^2}{L^2}))$,
and unit magnetic charge.

The general solutions are constructed numerically.
However, one can write also an approximate expression close to the horizon and for large-$r$;
the first terms in a near-horizon power series expansion in  $r-r_H$ read
\begin{eqnarray}
\nonumber
&&
m(r)= \frac{r_H(r_H^2+L^2)}{2L^2}+\alpha^2\frac{(1-w_H^2)^2}{2r_H^2}(r-r_H)+\dots,
\\
\label{rh}
&&
\sigma(r)=\sigma_H +\frac{2\sigma_H \alpha^2r_HL^4w_H^2(1-w_H^2)^2)}
{\left(3r_H^4+r_H^2L^2-\alpha^2 L^2(1-w_H^2)^2 \right)^2} (r-r_H)+\dots,
\\
\nonumber 
&&
w(r)=w_H+\frac{r_HL^2 w_H(w_H^2-1)}{3r_H^4+r_H^2L^2-\alpha^2 L^2(1-w_H^2)^2}(r-r_H)+\dots~,
\end{eqnarray}
with $r_H$, $w_H $ and $\sigma_H$ input  (positive) parameters
 fixing the Hawking temperature and event horizon area of the solutions
(with the entropy $S=A_H/4G)$
\begin{eqnarray}
\label{THS}
T_H=\frac{1}{4\pi} \frac{\sigma_H\left(3r_H^4+r_H^2L^2-\alpha^2 L^2(1-w_H^2)^2 \right)}{r_H^3L^2},~~A_H=4\pi r_H^2.
\end{eqnarray}
A similar expression can also be written for $r\to \infty$ as a power series in $1/r$, with
\begin{eqnarray}
\label{inf}
m(r)=M-\frac{\alpha^2 (2J^2+L^2(1-w_0^2)^2)}{2L^2 r}+\dots ,~~
\sigma(r)=1-\frac{1}{2r^4}\alpha^2 J^2 +\dots,~~
w(r)=w_0-\frac{J}{r}+\dots,
\end{eqnarray}
with three free parameters: $M$, which fixes the ADM mass of the solutions,  
$w_0$ which gives the (SU(2)-valued) magnetic charge 
${\cal Q}_M=Q_M\frac{\tau_3}{2\hat g}$ (where $Q_M=1-w_0^2)$
and $J$--an order parameter which provides a measure of non-Abelianess.

Note also that the equations (\ref{eqs1}) 
possess two scaling symmetries: $(i)~\sigma\to \lambda \sigma$
and $(ii)~r\to \lambda r,~~m\to \lambda r,~~L\to\lambda L,~~\alpha \to\lambda \alpha$
(with $\lambda$ a positive scaling parameter). The symmetry   $(i)$ was used to set  
  $\sigma(r)\to 1$  as $r\to \infty$.
The symmetry $(ii)$ can be used 
to fix the value of the AdS radius $L$ or the value of the coupling constant $\alpha$;  
in this work we set $L=1$ and treat $\alpha$ as an input parameter 
 (also, to simplify the expression of some quantities, 
we set $G=1$).

Restricting to a canonical ensemble
(which is the natural one in the presence of a magnetic charge), 
we study solutions holding the temperature $T_H$ and the magnetic
charge $Q_M$ fixed. The associated thermodynamic
potential is the Helmholtz free energy
\begin{eqnarray}
F=M-T_H S.
\end{eqnarray}
 When different solutions exist at fixed $(T_H,Q_M)$,
the configuration with lowest $F$ is the one that is thermodynamically favoured.
 Also, the condition for (local) thermodynamic stability of BHs 
in the canonical ensemble is
 \begin{eqnarray}
C=\left (\frac{\partial M}{\partial T_H} \right )_{Q_M}=T_H\left (\frac{\partial S}{\partial T_H} \right )_{Q_M}>0.
\end{eqnarray}
 As recently found in \cite{Fan:2014ixa},
the generic hairy solutions satisfy 
 the first law of thermodynamics 
%with $w_0\neq 0$ 
%
 %
 \begin{eqnarray}
\label{1slaw}
dM=\frac{\sigma_H}{2r_H^2L^2} \left(3r_H^4+r_H^2L^2-\alpha^2 L^2(1-w_H^2)^2 \right)    dr_H+\frac{\alpha^2 w_0 J}{L^2} dw_0
=T_H dS+\Phi dQ_M,
\end{eqnarray}
 where $\Phi=-\alpha^2   J/(2w_0L^2)$ is the  magnetostatic potential.

The most remarkable feature 
of the AdS EYM BHs is perhaps that 
the value of the parameter $w_0$ in (\ref{inf}) 
($i.e.$ the value of the magnetic charge) 
is not fixed $apriori$.
That is, for given $\alpha$
and any horizon size (as set by the input parameter $r_H$),
solutions are found for intervals of $w_0$
and not only $w_0^2=1$, as for $\Lambda=0$ 
(note however, that the allowed range of $w_0$
decreases as  $\alpha$ increases 
\cite{Winstanley:1998sn},
\cite{Bjoraker:2000qd},
\cite{Hosotani:2001iz}).  

 %%%%%%%%%%%%%%%%%%%%%%%%%%%%%%%%%%%%%%%%%%%%%%%%%%%%%%%%%%%
 \setlength{\unitlength}{1cm}
 \begin{picture}(8,6) 
\put(-1,0.0){\epsfig{file=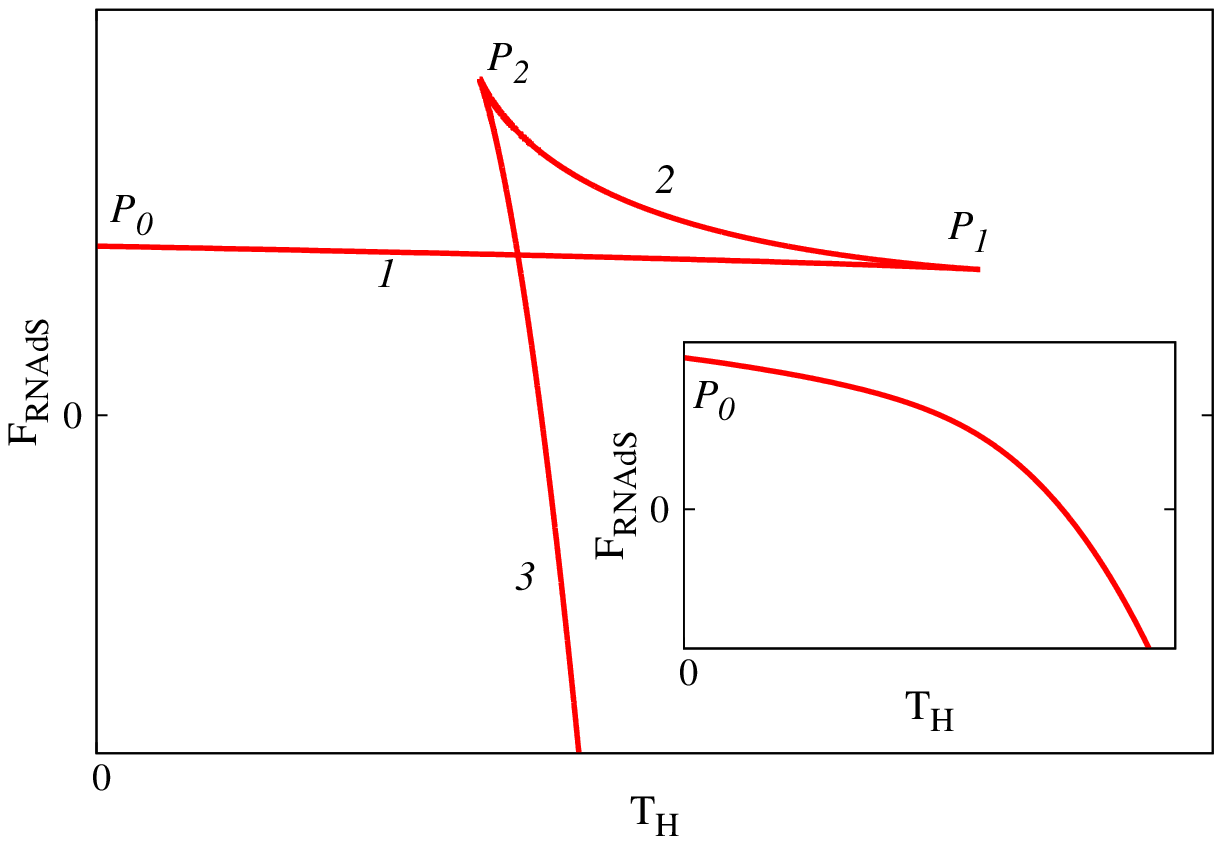,width=8.5cm}}
\put(8.2,0.0){\epsfig{file=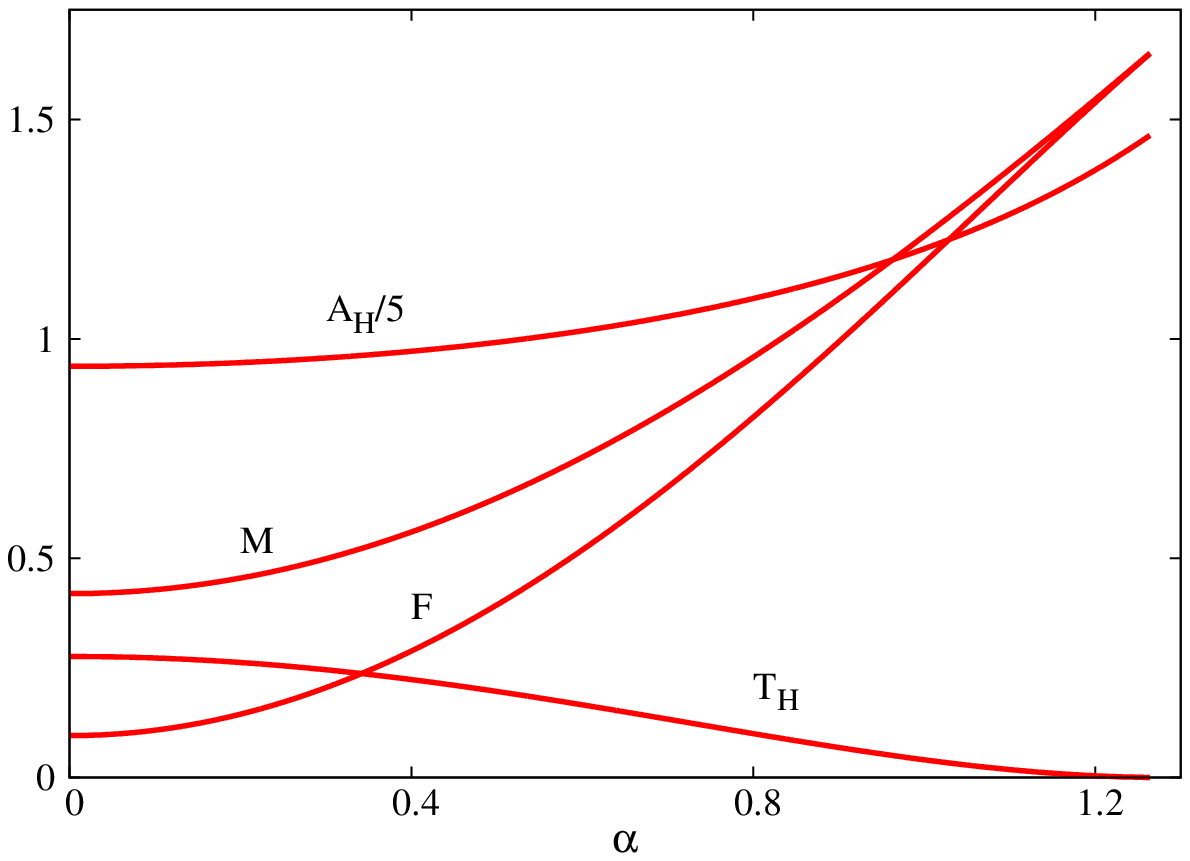,width=8.5cm}}
\end{picture}
\\
{\small {\bf Figure 1.} 
{\it Left:} The free energy is shown as a function of temperature for generic embedded Abelian 
Reissner-Nordstr\"om-AdS black holes with $0<\alpha<L/6$, in which case
 one notices the existence of three branches of solutions.
The generic picture for $\alpha>L/6$
is displayed in the inset, with the existence of a single branch of solutions.
{\it Right:} The mass, free energy, temperature and area of the unstable 
Reissner-Nordstr\"om-AdS black holes where a branch of non-Abelian solutions
emerges is shown as a function of $\alpha$.
 }
\vspace{0.5cm}
%%%%%%%%%%%%%%%%%%%%%%%%%%%%%%%%%%%%%%%%%%%%%%%%%%%%%%%%%%%%

Moreover, 
as for the better known case of asymptotically flat hairy BHs  \cite{Volkov:1989fi},
the solutions here are also indexed by the node number of the magnetic potential.

In this work we shall
consider nodeless  (for $w_0>0$)
and one node solutions (for $w_0<0$)
only.  
The
 configurations with higher number of nodes represent excited states
%whose energy is always greater than the energy of the
%corresponding nodeless configurations,  
and are therefore
 ignored in what follows.
Moreover, the solutions with nodes are unstable in linearized perturbation theory.
A detailed discussion  of these aspects 
together with a large set of references
can be found in \cite{Winstanley:2008ac}.

%%%%%%%%%%%%%%%%%%%%%%%%%%%%%%%%%%%%%%%%%%%%%%%%%%%%%%%%%%%%%%%%%%%%%%%%%%%%%%
\section{The thermodynamics of non-Abelian black holes }
%%%%%%%%%%%%%%%%%%%%%%%%%%%%%%%%%%%%%%%%%%%%%%%%%%%%%%%%%%%%%%%%%%%%%%%%%%%%%%

%%%%%%%%%%%%%%%%%%%%%%%%%%%%%%%%%%%%%%%%%%%%%%%%%%%%%%%%%%%%%%%%%%%%%%%%%%%%%%
\subsection{The  unit magnetic charge solutions }
%%%%%%%%%%%%%%%%%%%%%%%%%%%%%%%%%%%%%%%%%%%%%%%%%%%%%%%%%%%%%%%%%%%%%%%%%%%%%%

Among all sets of nA BHs, 
of particular interest are those solutions 
possessing a unit magnetic charge, 
$i.e.$
with 
$w_0=0$ in the far field expansion (\ref{inf}).
The existence of such configurations provides an
obvious violation of the no-hair conjecture, since two  distinct 
solutions are found for the same set of global charges.
That is,
apart from the gravitating Dirac mononopole
configuration (\ref{RNAdS}) with $w(r)\equiv 0$,
there is also an (intrinsic nA) configuration  
possessing a nontrivial profile of the magnetic potential $w(r)$.

%We introduced the following notation:
%\begin{eqnarray}
%x=6\frac{\alpha}{L},~~y=2\sqrt{3}\frac{\alpha}{L},~~
%q_{\pm}=\sqrt{1\pm \sqrt{1-(6\frac{\alpha}{L})^2}},~~p_{\pm}=\sqrt{ \sqrt{1+(2\sqrt{3}\frac{\alpha}{L})^2}\pm 1}.
%\end{eqnarray}

At this point it is instructive to briefly review the thermodynamics 
of the magnetically charged RNAdS solutions\footnote{The thermodynamics of RNAdS solution has been discussed from a slightly different perspective in 
\cite{Chamblin:1999tk}.
%\cite{Chamblin:1999hg},
for electrically charged BHs; 
however,
due to the existence of electric-magnetic duality in $d = 4$ Einstein-Maxwell theory, the results there apply to
magnetically charged
RNAdS solutions as well.} (\ref{RNAdS}).
The free-energy $vs.$ temperature behaviour of these solutions is summarized in Figure 1 (left).
For any $\alpha>0$,
a branch of RNAdS BHs emerges at
% $P_0\big (0, \frac{12\alpha^2+(-1+\sqrt{1+\frac{12\alpha^2}{L^2}})L^2}{3\sqrt{6}L\sqrt{-1+\sqrt{1+\frac{12\alpha^2}{L^2}}}} \big)$,
$P_0\big (0, \frac{L}{2\sqrt{3}}p_{-}(1+p_{+}^2) \big)$,
a point which corresponds to extremal BHs with a nonzero area 
%$A_H(T_H=0)= \frac{2\pi L^2}{3}(\sqrt{1+\frac{12\alpha^2}{L^2}}-1)$.
$A_H(T_H=0)= \frac{2\pi L^2}{3}p_{-}^2$ (where we note 
$
p_{\pm}=\sqrt{ \sqrt{1+(2\sqrt{3}\frac{\alpha}{L})^2}\pm 1}
$
and
$
q_{\pm}=\sqrt{1\pm \sqrt{1-(\frac{6\alpha}{L})^2}}
$
).
For $0<\alpha < L/6$,
this branch continues  to the point 
% $P_1\bigg(
% \frac{\sqrt{3}}{\pi L\sqrt{2}}  \frac{1-\sqrt{1-\frac{36\alpha^2}{L^2}}-\frac{12\alpha^2}{L^2}}{(1-\sqrt{1-\frac{36\alpha^2}%{L^2}})^{3/2}},
% \frac{L}{6\sqrt{6}}\frac{\frac{36\alpha^2}{L^2}+1-\sqrt{1-\frac{36\alpha^2}{L^2}}}{\sqrt{1-\sqrt{1-\frac{36\alpha^2}{L^2}}}}
% \bigg)$,
 $P_1\bigg(
 \frac{1}{\pi L\sqrt{3}} (q_{-}+\frac{1}{q_{-}}),
 \frac{L}{6\sqrt{6}}q_{-}(1+q_{+}^2)
 \bigg)$,
where a cusp occurs signaling a second branch of solutions extending backward in $T_H$, 
with an increasing $F$
%(the corresponding value of the mass at $P_1$ is $\frac{L\sqrt{2}}{3\sqrt{3}}\sqrt{1-\sqrt{1-\frac{36\alpha^2}{L^2}}}$).
(the corresponding value of the mass at $P_1$ is $\frac{L\sqrt{2}}{3\sqrt{3}}q_{-}^2$).
%

%%%%%%%%%%%%%%%%%%%%%%%%%%%%%%%%%%%%%%%%%%%%%%%%%%%%%%%%%%%%%%%%%%%
\setlength{\unitlength}{1cm}
\begin{picture}(15,20.85)
\put(-0.25,10){\epsfig{file=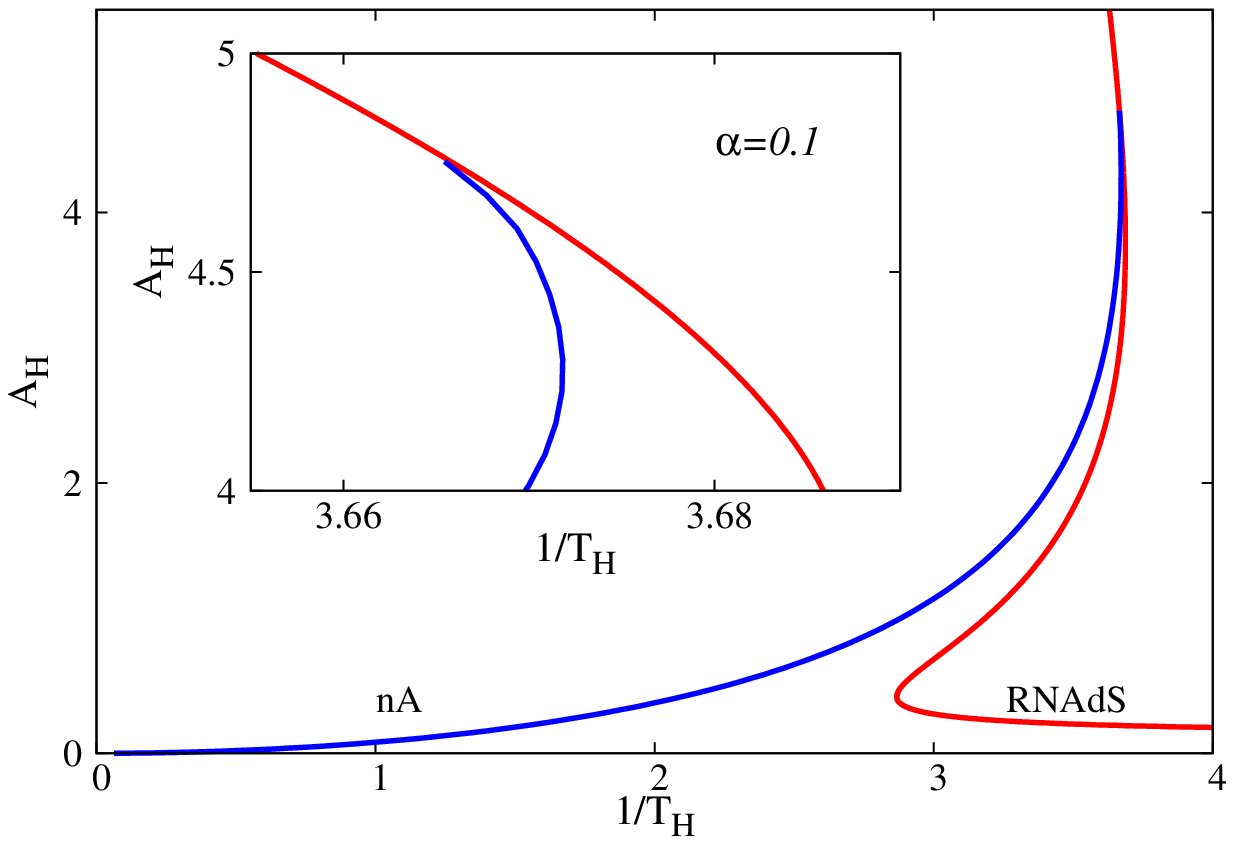,width=7.5cm}}
\put(8,10){\epsfig{file=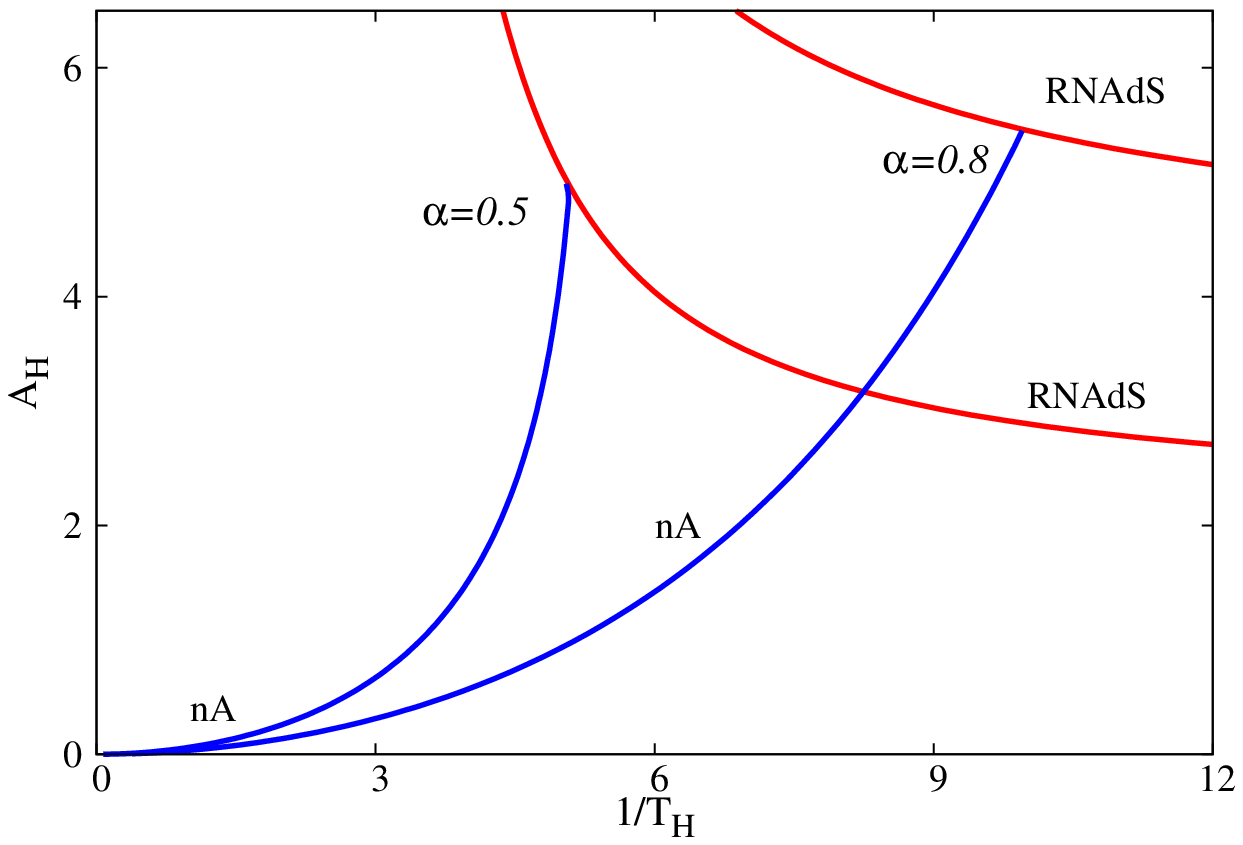,width=7.5cm}}
\put(-0.5,16){\epsfig{file=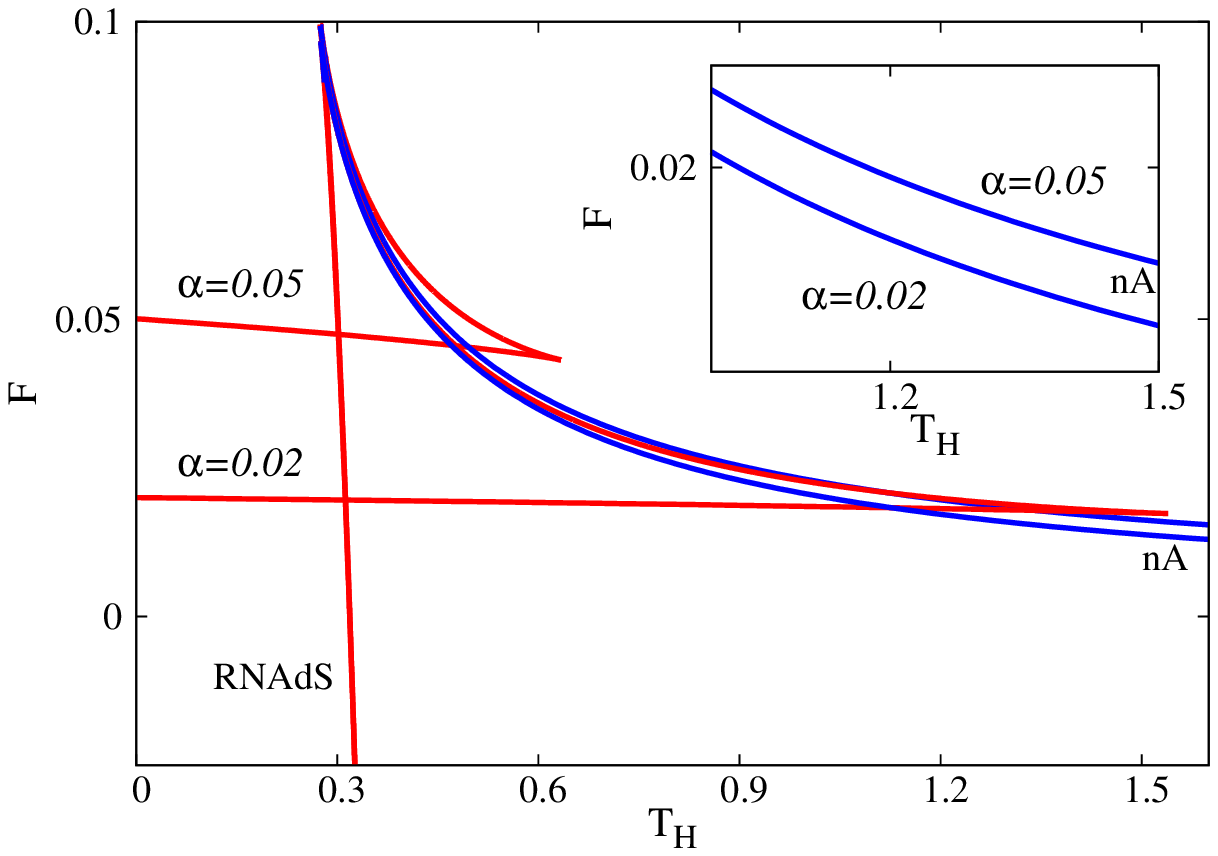,width=7.5cm}}
\put(8,16){\epsfig{file=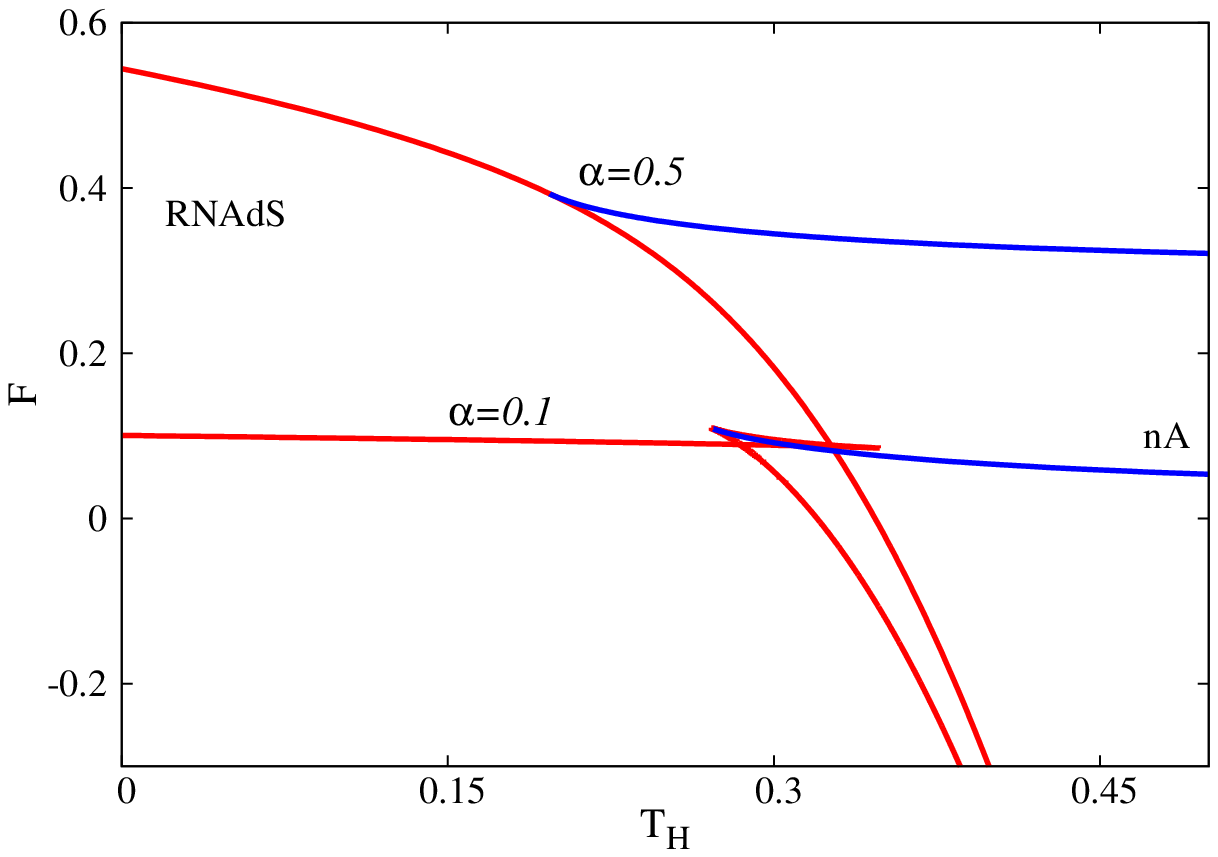,width=7.5cm}}
\end{picture}
\vspace*{-9.5cm}
\\
{\small {\bf Figure 2.} 
The  free energy and the horizon area are shown as a function of temperature for 
 embedded Abelian (red curves) and non-Abelian  (blue curves) solutions with unit magnetic
charge and several values of $\alpha$.}
\vspace{0.5cm}
%%%%%%%%%%%%%%%%%%%%%%%%%%%%%%%%%%%%%%%%%%%%%%%%%%%%%%%%%%%%%%%

This secondary branch ends at  
% $P_2\bigg(
% \frac{\sqrt{3}}{\pi L\sqrt{2}}  \frac{1+\sqrt{1-\frac{36\alpha^2}{L^2}}-\frac{12\alpha^2}{L^2}}{(1+\sqrt{1-\frac{36\alpha^2}%{L^2}})^{3/2}},
% \frac{L}{6\sqrt{6}}\frac{\frac{36\alpha^2}{L^2}+1+\sqrt{1-\frac{36\alpha^2}{L^2}}}{\sqrt{1+\sqrt{1-\frac{36\alpha^2}{L^2}}}}
% \bigg)$,
 $P_2\bigg(
 \frac{1}{\pi L\sqrt{3}} (q_{+}+\frac{1}{q_{+}}),
 \frac{L}{6\sqrt{6}}q_{+}(1+q_{-}^2)
 \bigg)$,
where another cusp is found, with the occurrence of a third branch of solutions
%(the corresponding value of the mass at $P_2$ is $\frac{L\sqrt{2}}{3\sqrt{3}}\sqrt{1+\sqrt{1-\frac{36\alpha^2}{L^2}}}$). 
(the corresponding value of the mass at $P_2$ is $\frac{L\sqrt{2}}{3\sqrt{3}}q_{+}^2$). 
This branch interesects the first one at some $T_H^{(c)}$ (with the occurrence of a ``swallowtail`` shape) 
and extends to arbitrarily large $T_H$
(note that the free energy vanishes for 
%$T_H=\frac{\sqrt{2}}{\pi L}\frac{1+\frac{4\alpha^2}{L^2}+\sqrt{1+\frac{12\alpha^2}{L^2}}}{(1+\sqrt{1+\frac{12\alpha^2}{L^2}})^{3/2}}$ 
$T_H=\frac{\sqrt{2}}{\pi L}(p_{+}+\frac{1}{p_{+}})$ 
%(a point where $M=\frac{L}{\sqrt{2}}\frac{1+\frac{4\alpha^2}{L^2}+\sqrt{1+\frac{12\alpha^2}{L^2}}}{\sqrt{1+\sqrt{1+\frac{12\alpha^2}%%{L^2}}}}$),
(a point where $M=\frac{L}{3\sqrt{2}}p_{+}(1+p_{+}^2)$)
and becomes negative for larger values of $T_H$.
%

%As discussed in \cite{Chamblin:1999tk},
The first branch solutions dominate
the thermodynamic ensemble for $T_H<T_H^{(c)}$;
however, for higher temperatures, the free energy is minimized by the 
solutions on the third branch.
However, 
the relative size of the second branch decreases with $\alpha$, 
the points $P_1$ and $P_2$ coinciding for $\alpha=L/6$, where the second branch of solutions disappears.
Then, for  $\alpha>L/6$ only one single branch of solutions is found,  see the inset in Figure 1 (left) 
(from there, it is also clear that, for any $\alpha$, 
at low enough temperatures there can only be one solution).

The picture valid for the  nA solutions is different.
First, unit magnetic charge solutions are found 
for 
$0<\alpha<\alpha_{max} $ only, 
with $\alpha_{max}\simeq 1.264$.
Another striking feature is the existence  of a 
soliton limit of the BHs,\footnote{An exact particle-like solution with $w_0=0$ is known for   $\alpha=0$ 
($i.e.$ the probe limit--YM fields in a fixed AdS background),
and has a magnetic potential $w(r)=1/\sqrt{1+\frac{r^2}{L^2}} $    \cite{BoutalebJoutei:1979va}.
Moreover, one can write a perturbative solitonic solution (with unit magnetic charge)
of the EYM eqs. (\ref{eqs1})
by considering a perturbative expansion in $\alpha^2$
around the global AdS background.
For example, the first order corrected metric functions are
$m(r)=\frac{3\alpha^2}{4}\left(\frac{1}{L} \arctan(\frac{r}{L})-\frac{r}{ L^2+r^2 } \right)$,
$\sigma(r)=1-\alpha^2 \frac{L^2}{2(L^2+r^2)^2}$.
 
This solution captures
some of the basic features of the general
nonperturbative configuration
and can also be extended  to higher orders in $\alpha$;
however, so far we could not identify a general pattern;
moreover,
the expressions for the functions become increasingly complicated.
} 
which is approached as $r_H\to 0$,  where $T_H\to \infty$, $A_H\to 0$
and $F\to M>0$.

\vspace{0.5cm}
 %%%%%%%%%%%%%%%%%%%%%%%%%%%%%%%%%%%%%%%%%%%%%%%%%%%%%%%%%%%
 \setlength{\unitlength}{1cm}
\begin{picture}(8,6) 
\put(-1,0.0){\epsfig{file=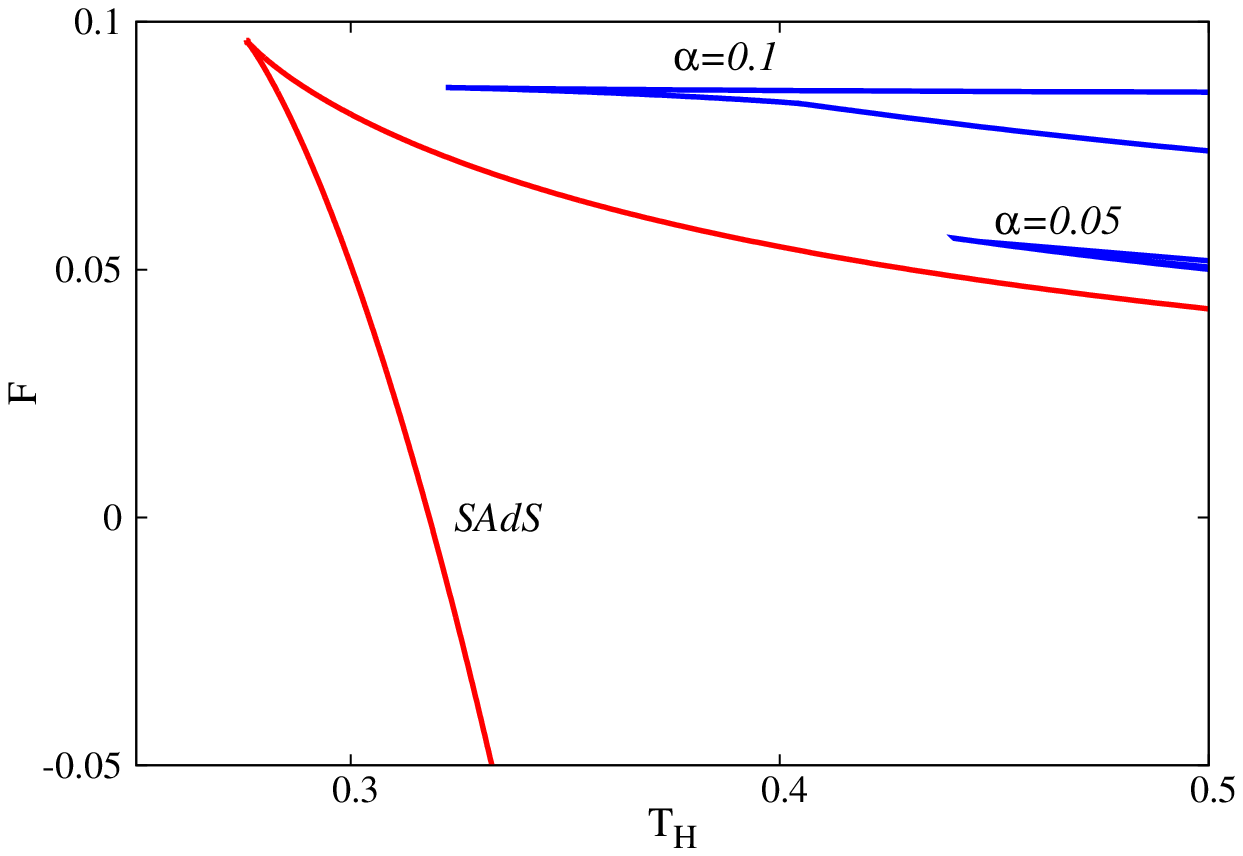,width=8.5cm}}
\put(8.2,0.0){\epsfig{file=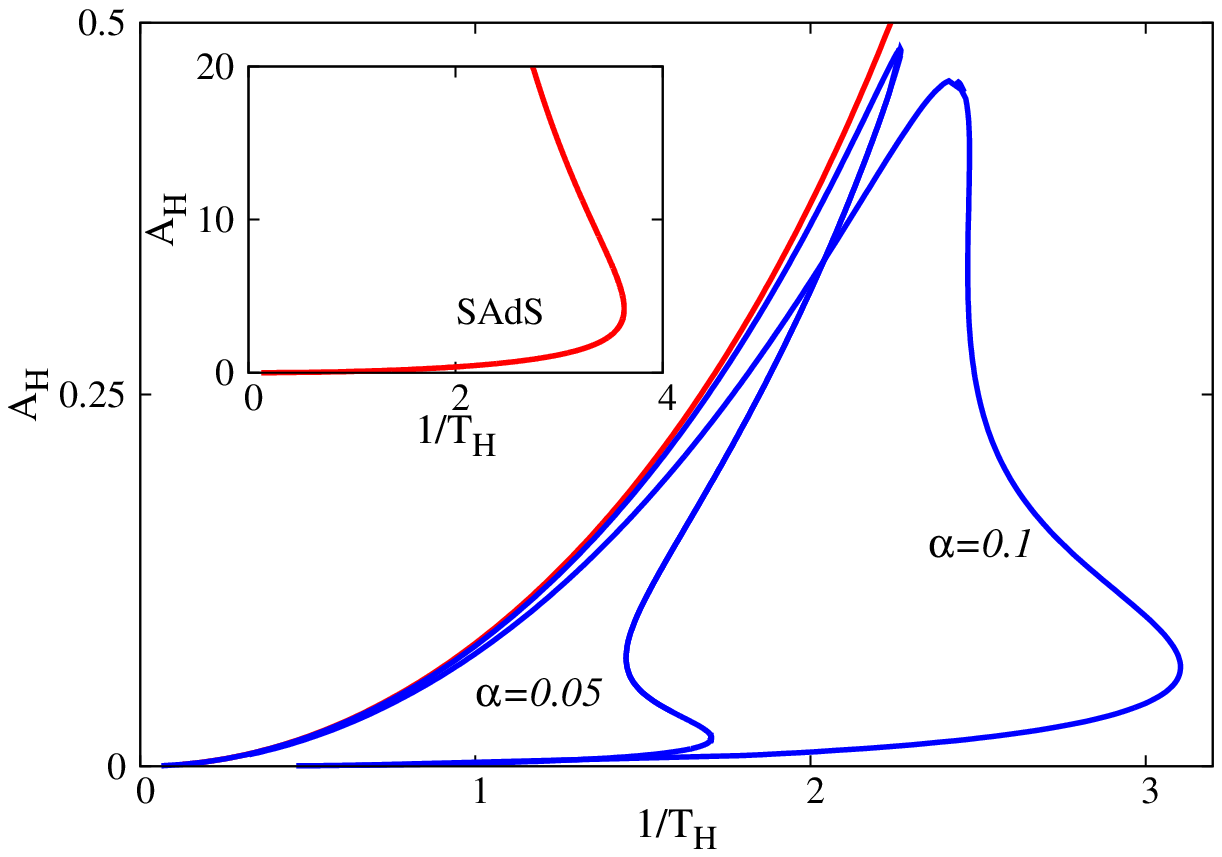,width=8.5cm}}
\end{picture}
\\
{\small {\bf Figure 3.} 
The  free energy and the horizon area are shown as a function of temperature for 
the vacuum Schwarzschild-AdS black holes (red curves) and non-Abelian (blue curves) solutions with vanishing
magnetic
charge and  two values of $\alpha$.
}
\vspace{0.5cm}
%%%%%%%%%%%%%%%%%%%%%%%%%%%%%%%%%%%%%%%%%%%%%%%%%%%%%%%%%%%%

 %%%%%%%%%%%%%%%%%%%%%%%%%%%%%%%%%%%%%%%%%%%%%%%%%%%%%%%%%%%
 \setlength{\unitlength}{1cm}
\begin{picture}(8,6) 
\put(-1,0.0){\epsfig{file=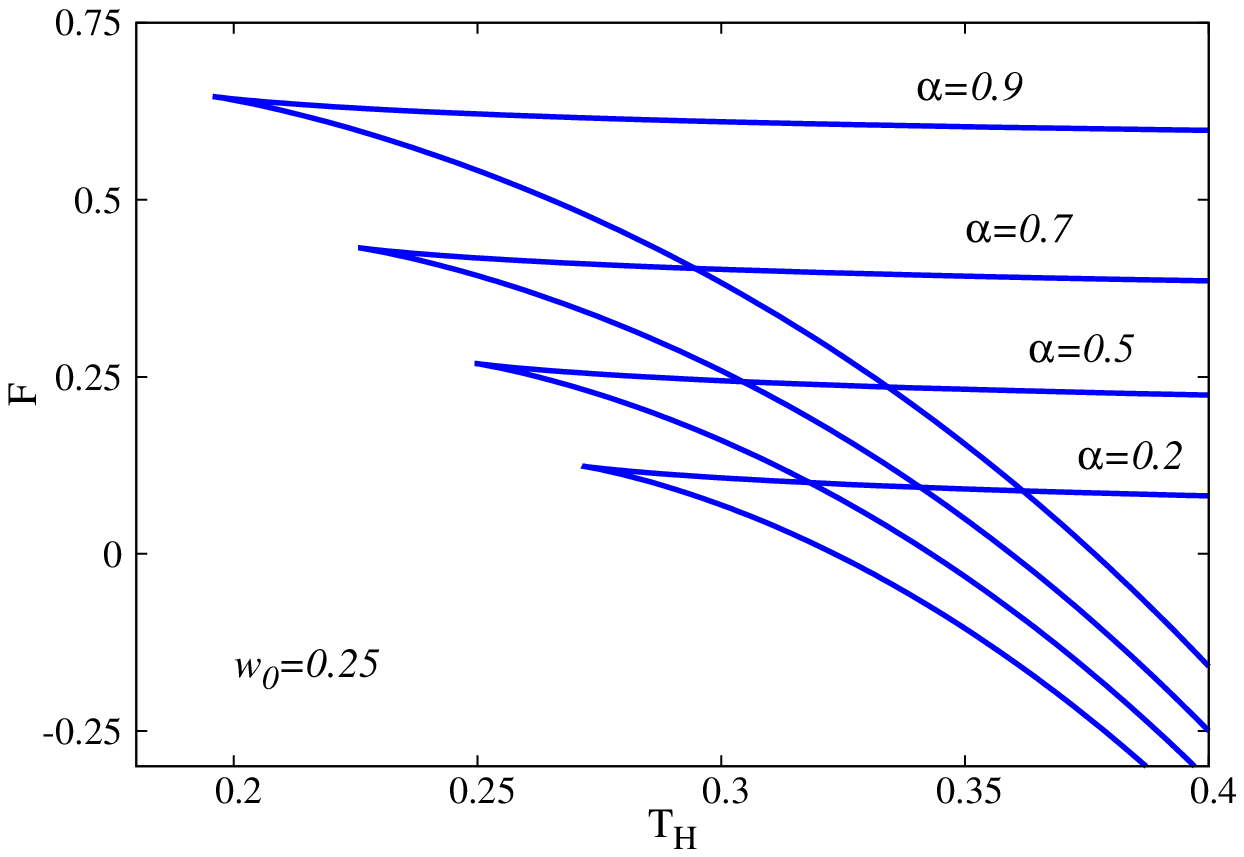,width=8.5cm}}
\put(8.2,0.0){\epsfig{file=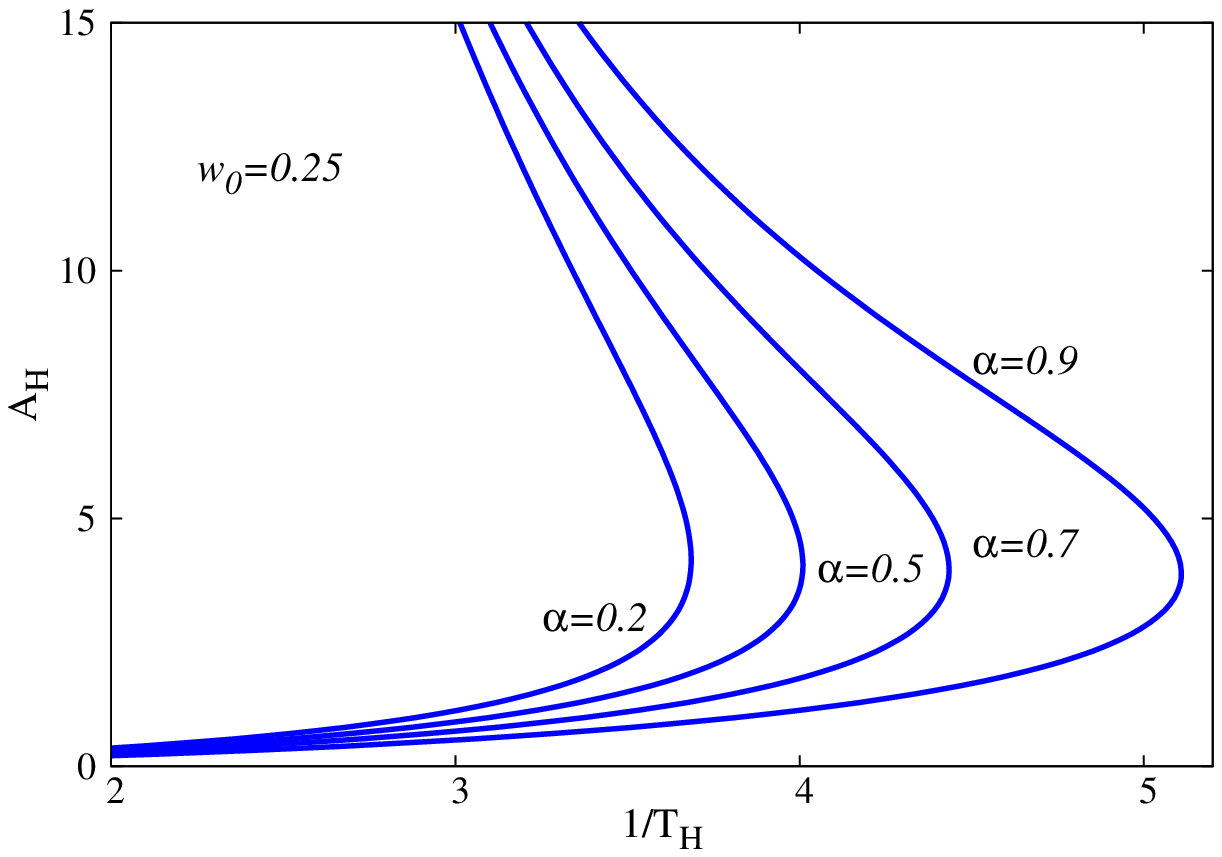,width=8.5cm}}
\end{picture}
\\
{\small {\bf Figure 4.} 
The  free energy and the horizon area are shown as a function of temperature for 
 generic non-Abelian black solutions solutions with a noninteger
magnetic charge and several values of $\alpha$.
}
\vspace{0.5cm}
%%%%%%%%%%%%%%%%%%%%%%%%%%%%%%%%%%%%%%%%%%%%%%%%%%%%%%%%%%%%
 
Moreover,
we notice that the nA solutions exist above a minimal value of $T_H>0$ only. 
In particular, no extremal BHs with nA hair is found,
 which agrees with the recent results in \cite{Li:2013gca}.
 %Also, for fixed $T_H$ and a given value of the mass parameter,
% the entropy is maximized by the third branch of Abelian solution.
% (provided
%it exist for that temperature).

Also, as seen in Figure 2, for a given temperature,
the free energy of all solutions is minimized by a RNAdS solution.
For $\alpha<L/6$, this RNAdS solution
 is located on the first branch (for $T<T_H^{(c)}$)  or on the third branch
 (for higher temperatures).
Thus we conclude that all unit magnetic
 charge
nA solutions are globally thermodynamically unstable (despite the fact that
their specific heat can be positive for some range of $r_H$).

Remarkably, one finds that, as the minimal value of  $T_H$ is approached,
 the branch of nA solutions
joins smoothly a critical RNAdS solution with the same value of $\alpha$, 
such that the magnetic
gauge function 
$w(r)$ becomes identically zero.
This bifurcation can also be viewed as indicating  that
the  embedded RNAdS BH presents an
instability with respect to static nA perturbations, for a critical value of the horizon radius.
This instability can be studied within the same Ansatz (\ref{YM-ansatz}),
 by considering values of the magnetic gauge potential $w(r)$
close to zero everywhere, $w(r) =  \epsilon W(r)$,
 and a fixed RNAdS background. 
The perturbation $W(r)$ starts from some
nonzero value at the horizon and vanishes at infinity, being a solution of the linear equation
\begin{eqnarray}
(N(r)W'(r))'+\frac{W(r)}{r^2}=0,
\end{eqnarray}
with $N(r)$ given by (\ref{RNAdS}).
Although the above equation  does not appear to be solvable in terms of known functions, one can
write again an approximate solution near the horizon and at infinity.
The general solution  
 is constructed numerically.
Given
$0<\alpha<\alpha_{max} $,  
this reduces to finding the critical value of $r_H$
such that the function $W(r)$ has the  proper behaviour.
Some parameters of the critical RNAdS solutions are shown in Figure 1 (right).
%(note that an extremal RNAdS background is approached as  $\alpha \to \alpha_{max} $).
For $0<\alpha<L/6$,
the unstable RNAdS solutions are located on the  second and third branches 
in a small region around the point $P_2$ (see Figure 1 (left)).
Further increasing $\alpha$, 
the unstable solutions move to smaller temperatures,
the point $P_0$ being approached for   $\alpha \to \alpha_{max} $,
with $W(r)\equiv 0$ in that limit.

Also, an analytic estimate for the critical horizon radius of the RNAdS BHs, $r_H(\alpha)$,
can be found by matching at some intermediate point 
the expansion of
$W(r)$ 
(and its first derivative) at the horizon, to that at infinity. 
Surprisingly, it turns
out that the simple expression $r_H=\frac{\sqrt{\alpha L}}{3^{1/4}}$
provides a good approximation for most of the numerical data (typically with several percent error).

%%%%%%%%%%%%%%%%%%%%%%%%%%%%%%%%%%%%%%%%%%%%%%%%%%%%%%%%%%%%%%%%%%%%%%%%%%%%%%
\subsection{Solutions with a vanishing magnetic charge }
%%%%%%%%%%%%%%%%%%%%%%%%%%%%%%%%%%%%%%%%%%%%%%%%%%%%%%%%%%%%%%%%%%%%%%%%%%%%%%

Another case of interest corresponds to 
solutions with $w_0^2=1$ in the far field asymptotics (\ref{inf}) 
($i.e.$ with a vanishing magnetic charge)
which possess, however, a non-vanishing magnetic field in the bulk\footnote{
These are the natural AdS counterparts 
of the well-known asympotically flat solutions in  \cite{Volkov:1989fi},
which necessarily have $w_0^2=1$.
Also, note that
these are unstable configurations with respect to small perturbations,
the magnetic potential possessing at least one node.
}.

The picture we have found is rather different in this case.
First, these hairy BHs do not emerge as perturbations\footnote{That is, one can show that the linearized YM equation in (\ref{eqs1})
with $w(r)=-1+\epsilon W_1(r)$
in a SAdS background does not possess solutions which are finite at $r=r_H$
and vanish asymptotically.}
 of the SAdS solutions (\ref{SAdS}).
Instead of  that, one finds two branches of solutions,
which, in a $F(T_H)$ diagram form a cusp for some minimal value of $T_H>0$.
As seen in Figure 3, 
this minimal value of $T_H$ decreases with $\alpha$
 (the well-known $F(T_H)$ diagram for the SAdS black holes is also shown there;
note that the vacuum solutions do not possess a dependence on $\alpha$).
 Moreover, the large SAdS solution is 
always thermodynamically favoured, minimizing the free energy.

Here it is interesting to consider the $\Lambda \to 0$
limit of these EYM BH solutions.
Then, by using the data in \cite{Volkov:1989fi}, one can easily verify that the
difference between the free energy of a hairy BH and the Schwarzschild solution with the same
temperature is always positive. 
The asymptotically flat EYM solutions are known to be unstable in perturbation theory
\cite{Volkov:1998cc}.
Then their thermodynamics also suggests that they should decay to a Schwarzschild vacuum BH.

%%%%%%%%%%%%%%%%%%%%%%%%%%%%%%%%%%%%%%%%%%%%%%%%%%%%%%%%%%%%%%%%%%%%%%%%%%%%%%
\subsection{The general case}
%%%%%%%%%%%%%%%%%%%%%%%%%%%%%%%%%%%%%%%%%%%%%%%%%%%%%%%%%%%%%%%%%%%%%%%%%%%%%%
The generic solutions have $w_0^2\neq 1$ and $w_0\neq 0$,
 possessing a nonvanishing (and noninteger) magnetic charge\footnote{Strictly speaking,
 these EYM BHs do not violate the no hair conjecture (we recall that 
 the RNAdS solution possesses a unit magnetic charge only, when viewed as solution of the EYM theory). 
 }.
 As shown in Figure 4, the thermodynamics of the generic solutions resembles
 that of the SAdS BHs, with the existence of two branches of solutions.
However, in this case, the small BHs branch
emerges from a solitonic solution  with a vanishing horizon area and a nonzero
mass, extending backwards in $T_H$.
Then a cusp occurs for a minimal temperature, with the emergence
of a branch of large BHs.
The free energy is minimized by the secondary branch solutions,
which possess also a positive specific heat, $C>0$.
Also, note that a similar picture has been found for other values of $w_0$
apart from the one in Figure 4.

  %%%%%%%%%%%%%%%%%%%%%%%%%%%%%%%%%%%%%%%%%%%%%%%%%%%%%%%%%%%
 \setlength{\unitlength}{1cm}
\begin{picture}(8,6) 
\put(-1,0.0){\epsfig{file=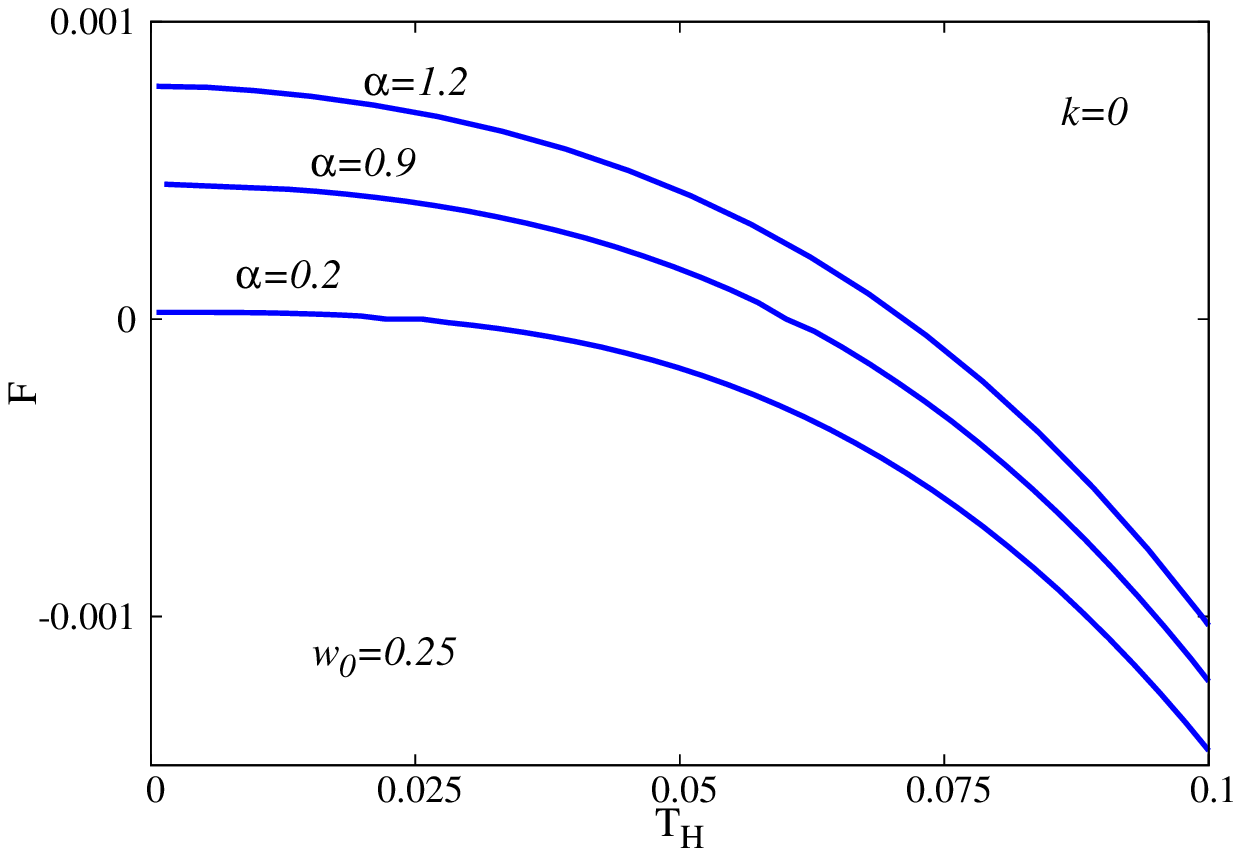,width=8.5cm}}
\put(8.2,0.0){\epsfig{file=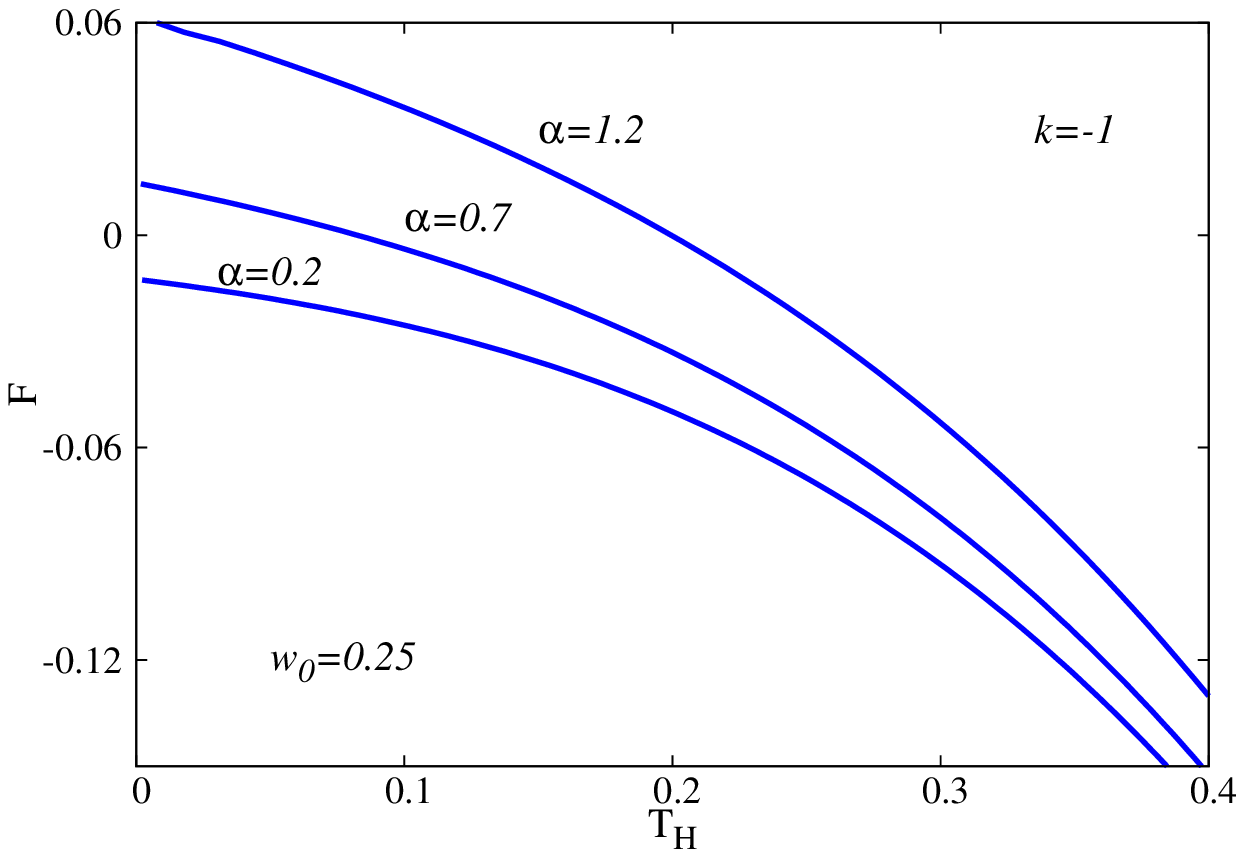,width=8.5cm}}
\end{picture}
\\
{\small {\bf Figure 5.} 
The  free energy density  is shown as a function of temperature for 
EYM black holes solutions with planar $(k=0)$ and hyperbolic $(k=-1)$ horizon topology and  several values of $\alpha$.
}
\vspace{0.5cm}
%%%%%%%%%%%%%%%%%%%%%%%%%%%%%%%%%%%%%%%%%%%%%%%%%%%%%%%%%%%% 

%%%%%%%%%%%%%%%%%%%%%%%%%%%%%%%%%%%%%%%%%%%%%%%%%%%%%%%%%%%%%%%%%%%%%%%%%%%%%%
\section{Further remarks}
%%%%%%%%%%%%%%%%%%%%%%%%%%%%%%%%%%%%%%%%%%%%%%%%%%%%%%%%%%%%%%%%%%%%%%%%%%%%%%
It is well known that the RNAdS BH solutions possess a variety of interesting thermodynamical
features,
see $e.g.$
\cite{Chamblin:1999tk}.
%\cite{Chamblin:1999hg}.
As expected, we have found that enlarging the gauge group of the matter fields to SU(2) 
leads to an even more complicated picture. 
For example, as seen already in other cases  
when the Einstein-Maxwell system is considered as part of a larger theory 
(see $e.g.$ \cite{Gubser:2008px}),
 we have found that the (unit magnetic charge) RNAdS BH can become unstable to forming hair at low temperatures.
This results in new branches of hairy solutions of the
larger theory\footnote{A related example is the case of asymptotically flat, magnetically
charged  RN BHs embedded in Einstein-Yang-Mills-Higgs theory, which develop
nA hair for some range of the parameters \cite{Lee:1991qs}.
For $\Lambda<0$, it is the asymptotic structure of spacetime which effectively replaces the Higgs field.}.
However, our results show  that 
the hairy BHs are never thermodynamically favoured over the full set of RNAdS solutions.

There are various possible extensions of the results discussed above.
First, our preliminary
results indicate the existence of static axially symmetric BH solutions with nA hair.
These solutions are the counterparts 
 of the 
particle-like EYM configurations discussed in \cite{Kichakova:2014fta}
being constructed by using a similar approach,
and possess an event horizon of spherical topology,
which, however, is not a round sphere.
They are found for an axially symmetric 
generalization of the magnetic YM ansatz (\ref{YM-ansatz})
with two positive discrete parameters $n$ and $m$ (which are the azimuthal and polar winding numbers, respectively),
the spherically symmetric ansatz (\ref{YM-ansatz}) being recovered for $n=m=1$.
These BHs possess as well a continuous parameter $w_0$ (which is a generalization of the constant $w_0$
 entering the asymptotics (\ref{inf})),
 which fixes the magnetic charge of the solutions  \cite{Kichakova:2014fta}.

Again, of particular interest are $(i)$ the configurations 
whose far field asymptotics of the YM fields describe a gauge transformed charge$-n$ Abelian multimonopole,
and $(ii)$  the  configurations  with a vanishing net magnetic flux.
The axially symmetric solutions we have constructed so far in a more systematic 
way, have $m=1$ and $n=2,3,4$ and thus represent deformations of the configurations in Section 2, sharing
their basic properties.
In particular, as expected, we have found  that the charge-$n$ RNAdS solution
is always thermodynamically favoured over the nA ones.
We hope to report elsewhere on these aspects.

A rather different picture is found when considering {\it 'topological black hole'}
generalizations of the solutions in Section 2.
In this case, the
two-sphere in the metric ansatz (\ref{metric-ansatz})
is replaced by a two-dimensional space  of negative or vanishing curvature\footnote{However, 
such configurations possess the same amount of symmetry
 for any topology of the horizon.}  
(see $e.g.$ \cite{Lemos:1994xp} for vacuum solutions),
the corresponding metric ansatz being
%These solutions are found for a direct generalization of (\ref{metric-ansatz}),
\begin{eqnarray}
\nonumber
ds^2=\frac{dr^2}{N(r)}+r^2(d\theta^2+f_k^2(\theta) d\phi^2)-\sigma^2(r)N(r) dt^2,
\end{eqnarray}
where $N(r)=k-\frac{2m(r)}{r}+\frac{r^2}{L^2}$, with
$k$   a discrete parameter  which fixes the topology of the horizon.
For $k=1$, $f_1=\sin \theta$ and the metric ansatz (\ref{metric-ansatz}) is recovered;
the solutions with $k=0$ have $f_0= \theta$ and are planar BHs,
approaching
 asymptotically a Poincar\'e patch of the AdS spacetime; finally, the solutions with $k=-1$
possess a hyperbolic horizon, with $f_{-1}=\sinh \theta$.
The corresponding general-$k$ YM ansatz has been displayed in \cite{VanderBij:2001ia} and  
looks very similar to (\ref{YM-ansatz}), with
\begin{eqnarray}
\nonumber
A=\frac{1}{2\hat g} \big [
w(r) \tau_1 d\theta+
(
\frac{df_k(\theta)}{d\theta} \tau_3 + f_k(\theta) w(r)\tau_2 
)
d\phi
\big ].
\end{eqnarray}
 The equations of the model are still given by 
 (\ref{eqs1}), 
with $V(w)=k-w^2(r)$ in the general case.

A study of the general-$k$ case shows that the EYM BHs 
with a spherical event horizon topology ($k=1$) are special.
First, a soliton limit of the BHs exists only in this case \cite{VanderBij:2001ia};
also, the embedded Abelian solution can possess an instability for $k=1$ only.
Moreover,  the EYM
BHs with a planar or hyperbolic horizon topology
are intrinsic nA, since they cannot approach
asymptotically a vacuum SAdS or a RNAdS configuration
(note that the magnetic gauge potential $w(r)$
is a nodeless function for $k\neq 1$).
 In fact, an
embedded Abelian solution exists for $k=\pm 1$ only,
$i.e.$
$w(r)\equiv 0$ implies 
$m(r)=M- {\alpha^2 k^2}/{(2r)}$,
 $\sigma(r)=1$,
which is the (planar) SAdS BH for $k=0$.
Also, no reasonable YM vacuum state exists for $k=-1$.

As shown in Figure 5, in strong constrast to the $k=1$ case, 
the EYM BHs with a  non-spherical
horizon topology 
exhibit a single branch of solutions\footnote{For the data in Figure 5, 
we set to one the area of the $(\theta,\phi)$-sector of the metric.}.
Moreover, the $k=0,-1$ solutions are (locally) thermally stable, since $C>0$ (similar results 
have been found for other values of the magnetic charge, as fixed by the parameter $w_0$
in the asymptotic expansion (\ref{inf})).
 
Finally, it would be interesting to extend the analysis in this work
to BH dyons, $i.e.$ solutions 
featuring both magnetic and electric nA fields.
The study of a special class of such EYM black brane solutions ($i.e.$ with a planar horizon, $k=0$)
has led
to the discovery of holographic superconductors, describing condensed
phases of strongly coupled, planar, gauge theories in $d=3$ dimensions 
\cite{Gubser:2008zu},
\cite{Gubser:2008wv}
(note that in this case the magnetic field vanishes on the boundary,
such that the hairy solutions share the asymptotics
with the electrically charged RNAdS black branes). 
We expect the EYM solutions with a spherical or hyperbolic event horizon topology
to exhibit a similar pattern.

 \vspace{1.cm} 
 
{\bf\large Acknowledgements} \\ 
E.R. would like to thank Cristian Stelea for useful 
comments on an early version of this paper.
We acknowledge  support by the DFG Research
Training Group 1620 "Models of Gravity".
The work of 
E.R. is supported by the FCT-IF programme and the
CIDMA strategic project  UID/MAT/04106/2013.
Y.S.  acknowledges support from 
 A. von Humboldt Foundation
in the framework of the Institute Linkage Programm and the JINR
Heisenberg-Landau Programm.

%%%%%%%%%%%%%%%%%%%%%%%%%%%%%%%%%%%%%%%%%%%%%%%%%%%%%%%%%%%%%%%%%%%%%%%%%%%%%%%%%%%%%
 \begin{small}
 
%%%%%%%%%%%%%%%%%%%%%%%%%%%%%%%%%%%%%%%%%%%%%%%%%%%%%%%%%%%%%%%%%%%%%%%%%%%%%%
 \end{small}

\end{document}